\documentstyle[12pt,epsfig,amssymb]{article}
\begin{document}
\begin{flushright}
IHEP-01-18\\
\end{flushright}

\begin{center}

{\Large \bf Determination of the high-twist contribution to 
the structure function \boldmath{$xF^{\nu N}_3$}}

\vspace{1cm}
{\bf S.~I.~Alekhin, V.~B.~Anykeyev, A.~A.~Borisov,
N.~I.~Bozhko, S.~K.~Chernichenko, 
R.~M.~Fachrutdinov, V.~N.~Goryachev, S.~V.~Goryachev, M.~M.~Kirsanov,
A.~I.~Kononov, A.~S.~Kozhin, V.~V.~Lipajev,
A.~I.~Mukhin, Y.~I.~Salomatin,
Y.~M.~Sviridov, V.~L.~Tumakov, A.~S.~Vovenko} 

\vspace{0.1in}
{\baselineskip=14pt Institute for High Energy Physics, 142284 Protvino, Russia}

\vspace{0.1in}
{\bf Y.~A.~Batusov, S.~A.~Bunyatov, O.~Y.~Denisov,
M.~Y.~Kazarinov, O.~L.~Klimov, A.~V.~Krasnoperov,
Y.~A.~Nefedov, B.~A.~Popov, S.~N.~Prakhov, V.~I.~Snyatkov,
 V.~V.~Tereshchenko}

\vspace{0.1in}
{\baselineskip=14pt Joint Institute for Nuclear Research, Dubna, Russia} 

\begin{abstract}
We extract the high-twist contribution to the 
neutrino-nucleon structure function $xF_3^{(\nu+\overline{\nu})N}$ 
from the analysis of the data collected by the 
IHEP-JINR Neutrino Detector in the runs with the focused neutrino beams
at the IHEP 70 GeV proton synchrotron. The analysis is 
performed within the infrared renormalon (IRR) model of high twists
in order to extract the normalization parameter of the model. 
From the NLO QCD fit to our data
we obtained the value of the IRR model normalization parameter  
$\Lambda^2_{3}=0.69\pm0.37~({\rm exp})\pm0.16~({\rm theor})~{\rm GeV}^2$.
We also obtained 
$\Lambda^2_{3}=0.36\pm0.22~({\rm exp})\pm0.12~({\rm theor})~{\rm GeV}^2$
from a similar fit to the CCFR data. The average of both results is 
$\Lambda^2_{3}=0.44\pm0.19~({\rm exp})~{\rm GeV}^2$. 
\end{abstract}
\end{center}
{\bf PACS numbers:} 13.60.Hb,06.20.Jr,12.38.Bx\\
{\bf Keywords:} high twists, deep inelastic scattering
\newpage

{\bf 1.} Attempts to extract the high twist
(HT) contributions to the neutrino-nucleon
deep-inelastic scattering (DIS)
structure functions started many years ago \cite{barnett},
but have not lead to the ultimate answer up to now. 
The main difficulty in this study is that due to the linear rise of the total 
interaction cross-section with the incident neutrino 
energy $E_{\nu}$, the largest data samples have been collected in experiments
at relatively high neutrino energies 
$E_{\nu} > 50~{\rm GeV}$. The region of the small momentum transfered 
$Q$, which is most relevant for the study of the 
HT effects, is rather poorly populated by the data points coming from  
these experiments because of kinematical and/or methodical cuts.
At lower neutrino energies 
experiments with very high luminosity are necessary
to achieve the statistical precision in the structure function 
measurements sufficient for the quantitative estimation of the
HT contribution. 

The first results on the twist-4  
contribution to the neutrino-nucleon structure function 
$xF^{\nu N}_3$ extracted from the analysis of data collected in a single 
experiment were reported in Ref.~\cite{Varvell:1987qu}. The indication 
on a negative sign of this contribution given in this paper was later 
confirmed with a better precision in Ref.~\cite{Ammosov:1987gq}.
Nevertheless, the experimental errors were large in both cases which
did not allow for a conclusive 
comparison with the available theoretical models of HT. Later the CCFR
experiment at Fermilab collected large statistical 
data sample \cite{Seligman:1997mc}, which allowed for more precise 
determination of the twist-4 contribution to $xF_3^{\nu N}$ 
\cite{Kataev:1998nc,Alekhin:1999df}, but the precision is still poor.

The data of Ref.~\cite{Sidorov} from the IHEP-JINR Neutrino Detector 
can be used to improve our knowledge of the HT contribution to the DIS 
structure functions. This experiment used
a neutrino beam of relatively low energy 
$(E_{\nu}<30~{\rm GeV}$), but collected rather large statistics 
(5987 neutrino and 741 antineutrino
charged-current (CC) interactions). The lowest $Q^2$ is 0.55~GeV$^2$
and the HT contribution would clearly manifest itself 
as a power-like correction to the logarithmic-like 
leading twist (LT) dependence of the structure functions on $Q$. 
Meanwhile the $Q$ range spanned by the data is 
limited (maximal $Q^2$ is 20~GeV$^2$) and for this reason the
simultaneous determination of the 
power-like and logarithmic-like terms is difficult.
In the analysis of Ref.~\cite{Sidorov} we fixed the HT contribution 
as it was defined from other experiments 
and performed the next-to-leading-order (NLO) QCD analysis of 
our data in order to constraint the LT contribution.
The value of the strong coupling constant
$\alpha_s$, which mainly governs the $Q$ dependence of the LT term, 
was determined from this analysis as $\alpha(M_Z)=0.123^{+0.010}_{-0.013}$.

At the same time the $Q$ dependence 
of the LT contribution can be well constrained using the data from 
other experiments. The world average of $\alpha_s(M_Z)$
is known with the precision of about  
0.003 \cite{Bethke:2000ai} and one can perform the analysis 
complimentary to the one of Ref.~\cite{Sidorov}, i.e.,
fix the value of $\alpha_s(M_Z)$ at the world average and try to extract
the HT contribution from the data. An additional 
error estimated as variation of the results of the fit under variation 
of the world average of $\alpha_s(M_Z)$ within its uncertainty should be 
ascribed thereafter.
Meanwhile in most cases the total error in the HT contribution extracted 
using this approach would be less as compared with the results 
of the simultaneous fit of the HT and the LT terms to the data.
The reduction of the error depends
on the ratio of the error in $\alpha_s(M_Z)$ obtained in such 
simultaneous fit to the error in the world average. 
For our data this ratio is larger than 3 and for this reason 
we extracted the HT contribution value from these data 
using the fit with a fixed value of $\alpha_s$.

{\bf 2.} The analysis is based on the data 
collected with three independent exposures
of the IHEP-JINR Neutrino Detector~\cite{ND} to the wide-band neutrino
and antineutrino beams~\cite{beams} of Serpukhov U-70 accelerator.
The exposure to the antineutrino beam ($\overline\nu_\mu$-exposure)
was performed at the proton beam energy $E_p=70$~GeV, whereas the two
$\nu_\mu$-exposures were carried out at $E_p=70$~GeV and at $E_p=67$~GeV.
The energy of the selected $\nu_\mu$ ($\overline\nu_\mu$) CC
events was in the range of $6 < E_{\nu(\overline\nu)} < 28$~GeV.
The experimental set-up and the selection criteria of CC
neutrino and antineutrino interactions are discussed in Ref.~\cite{publ}.
The $F_2^{(\nu+\overline{\nu})N}$ and $x F_3^{(\nu+\overline{\nu})N}$ 
structure functions of nucleon have been measured as a function 
of $x$ averaged over all $Q^2$ permissible (the details of 
experimental procedures are described in Ref.~\cite{Sidorov}).

These data were analyzed in the NLO QCD approximation 
in the modified minimal-subtraction
($\overline{\rm MS}$) renormalization-factorization scheme.
The partons evolution code applied in this analysis was   
used earlier for the global fit of the parton distribution functions 
\cite{Alekhin:2000ch}. The boundary parton distributions 
were chosen in the form 
$$
xp_{\rm NS}(x,Q_0)=A_{\rm NS}x^{a_{\rm NS}}(1-x)^{b_{\rm NS}},~~~
xp_{\rm S}(x,Q_0)=A_{\rm S}(1-x)^{b_{\rm S}},
$$
\begin{equation}
xp_{\rm G}(x,Q_0)=A_{\rm G}(1-x)^{b_{\rm G}}
\label{eqn:pdfs}
\end{equation}
at $Q_0^2=0.5$~GeV$^2$, 
where indices NS, S, and G correspond to non-singlet, singlet, and 
gluon distributions, respectively. These distributions were substituted
in the expressions for the LT contributions to $F_{2,3}$ 
$$
xF_3^{(\nu+\overline{\nu})N,LT}(x,Q)
=\int_x^1 \frac{dz}{z}C_3^q(z,Q)p_{NS}(x/z,Q),
$$
$$
F_2^{(\nu+\overline{\nu})N,LT}(x,Q)
=\int_x^1 \frac{dz}{z}\left\{C_2^q(z,Q)
\left[p_{NS}(x/z,Q)+p_S(x/z,Q)\right]+\right.
$$
\begin{equation}
\left.+C_G(z,Q)p_{G}(x/z,Q)\right\}
\end{equation}
where $C(z,Q)$ are the perturbative QCD 
coefficient functions in the $\overline{\rm MS}$ scheme. The parameter
$A_{\rm NS}$ was calculated using the constraint
$\int_0^1 dx q_{\rm NS}=3$, and the parameter $A_{\rm G}$ -- from the 
momentum-conservation constraint, while other parameters 
of Eq.~(\ref{eqn:pdfs}) were fitted to the data.
The form of Eq.~(\ref{eqn:pdfs}) was checked to be flexible enough, 
i.e., its complication did not lead to the improvement of the fit.

The target mass (TM) corrections of $O(M^2/Q^2)$,
as they are given in Ref.~\cite{Georgi:1976ve}, 
were applied to the LT contribution.
The HT contribution was parameterized in the additive form and 
within the infrared renormalon model (IRR) \cite{Beneke:1999ui}. 
In this model the HT contribution 
is connected with the LT one by the known 
coefficient function and the only free parameter 
of the model is related to the total normalization. 
In particular, the HT contribution to $xF_3^{(\nu+\overline{\nu})N}$ reads
\cite{Stein:1996wk,Dasgupta:1996hh}
\begin{equation}
H_{3}(x,Q)=A_2'(F_3^{\nu N})
\int_x^1 \frac{dz}{z} C_3^{IRR}(z) p_{\rm NS}(x/z,Q),
\label{eqn:htf3}
\end{equation}
where $C_3^{IRR}$ is the IRR model coefficient function 
and $A_2'(F_3^{\nu N})$ defines the total normalization.
The HT contribution to $F_2^{(\nu+\overline{\nu})N}$ 
contains the non-singlet term 
similar to Eq.(\ref{eqn:htf3}) with the normalization parameter
$A_2'(F_2^{\nu N})$ and the respective coefficient function $C_2^{IRR}$.
In addition, the singlet and gluon terms   
calculated in Ref.~\cite{Stein:1998wr} also come to the 
expression for the HT contribution to $F_2$ as it is given by the IRR model,
but these terms are relevant for small $x$ only and for this reason we 
used the non-singlet approximation for the calculation 
of the IRR contribution to $F_2^{(\nu+\overline{\nu})N}$ 
as well as for $xF_3^{(\nu+\overline{\nu})N}$.
Following Ref.~\cite{Stein:1998wr},
we describe the general normalization of the HT contributions 
to $F_{2,3}$ by the parameters $\Lambda_{2,3}$, which 
are connected with the parameters $A_2'(F_{2,3}^{\nu N})$ 
by the relations
\begin{equation}
A_2'(F_{2,3}^{\nu N})=-\frac{2C_F}{\beta_0}\Lambda_{2,3}^2,
\label{eqn:a2lam}
\end{equation}
where $C_F=4/3$ and $\beta_0$ is the first coefficient of the 
QCD $\beta$-function. Both ways are completely 
equivalent if the number of active fermions in the expression for 
$\beta_0$ does not depend on $Q$. 
Meanwhile in order to provide self-consistency 
of the analysis, we changed $n_f$ in Eq.~(\ref{eqn:a2lam}) from 3 to 4 
at $Q$ equal to the $c$-quark
mass $m_c=1.5~{\rm GeV}$. For this reason the value of $A_2'$
depends on $Q$ in our case, although the numerical effect is 
inessential.  
The value of $\alpha_s(M_Z)$ was fixed at 0.118, 
which is close to the world average of Ref.~\cite{Bethke:2000ai}.
As one can see in Fig.1
the analyzed data are insensitive to the parameter $\Lambda_2$ 
and we fixed it at the value 
of 1~GeV$^2$ inspired by the results of Ref.~\cite{Dasgupta:1996hh} 
on the analysis of charged leptons DIS data.
The systematic errors on the data were accounted for 
in the covariance matrix approach, described in Ref.~\cite{Alekhin:2000es}.

\begin{table}[h]
\caption{The results of the fit of the IRR model to the data from 
different neutrino experiments. 
The value of $\chi^2$ over the number of data points (NDP) is given
in the last column.} 
\begin{center}
\begin{tabular}{|c|c|c|c|} \hline
Experiment & $\Lambda_3^2 [{\rm GeV}^2]$ 
& $\Lambda_2^2 [{\rm GeV}^2]$ & $\chi^2/{\rm NDP}$ \\ \hline
IHEP-JINR & $0.69\pm0.37$ & 1. & 3/12 \\ \hline
CCFR & $0.36\pm0.22$ & $0.91\pm0.77$ & 253/222 \\ \hline
\end{tabular}
\end{center}
\end{table}

\begin{figure}
\centerline{\epsfig{file=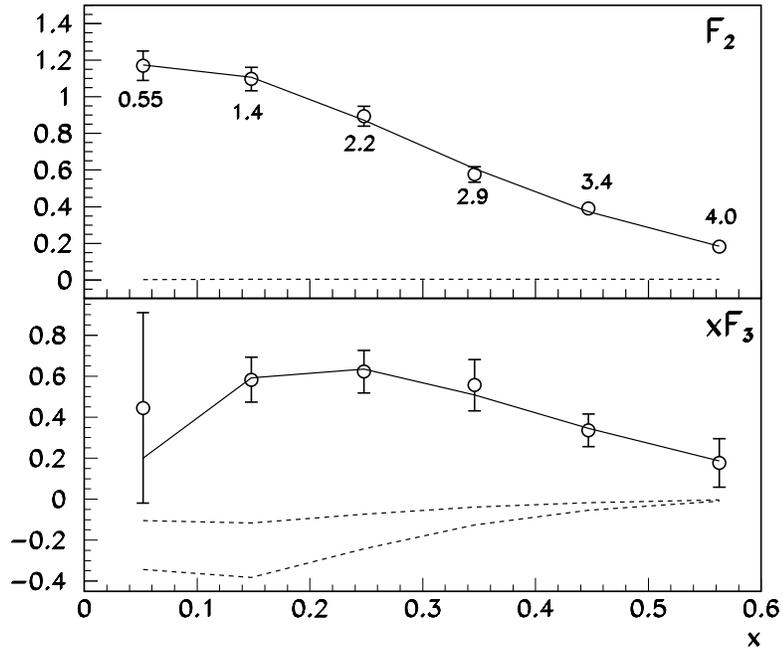,width=12cm,height=12cm}}
\caption{The $x$ dependence of the measured structure functions 
$F_2^{(\nu+\overline{\nu})N}$ (upper) and $xF_3^{(\nu+\overline{\nu})N}$ 
(lower). The average values of $Q^2$ (GeV$^2$)
for the $x$ bins are given in the upper plot. The full curves
give the result of the
LT+HT fit to the data, the dashed curves correspond to the $1\sigma$ bands
of the HT contributions obtained from the fit.}
\end{figure}

The results of the fit are given in Table~1 and in Fig.~1. The 
obtained contribution to $xF_3^{(\nu+\overline{\nu})N}$ 
is negative which supports the earlier observation 
of Refs.~\cite{Varvell:1987qu,Ammosov:1987gq}
 and is in agreement with the results of 
Refs.~\cite{Kataev:1998nc,Alekhin:1999df}.
The HT contribution to $F_2^{(\nu+\overline{\nu})N}$ is negligible 
in the whole region of $x$ spanned by the data.
The value of $\Lambda_3^2$ is determined from our data with the 
50\% accuracy. For the comparison, in the NLO QCD fit to the CCFR data
on the structure function $xF_3^{(\nu+\overline{\nu})N}$ the value 
$A_2'(F_3^{\nu N})=-0.12\pm0.05~{\rm GeV}^2$ 
was obtained in Ref.~\cite{Kataev:1998nc}. 
This estimate did not account for the
systematic errors in the data, while the estimate accounting for 
systematics is $A_2'(F_3^{\nu N})=-0.10\pm0.09~{\rm GeV}^2$ 
\cite{Alekhin:1999df}, i.e., 
our result is the most precise estimate of the IRR normalization 
parameter at the moment. 

The change of $\Lambda_3^2$ under variation of $\alpha_s(M_Z)$ by $\pm0.003$ is
0.055~GeV$^2$ and we consider this shift as a theoretical 
error in $\Lambda_3^2$. Another source of the theoretical error comes from 
the uncertainty due to the effect of the higher-order (HO) QCD corrections
to the LT term. These effects may be especially important  
for our study since the data at rather low $Q$ are involved in the analysis.
The HO corrections generally make the $Q$ dependence of 
the LT contribution steeper, and correspondingly lead to the decrease of the  
HT contribution. In order to estimate the
uncertainty due to neglected HO corrections, we repeated the fit with the 
QCD renormalization scale changed from the nominal value of $Q$ to 
$2Q$ (see Ref.~\cite{Martin:1991jd} for a 
detailed argumentation of this approach).
The obtained shift in the value of $\Lambda_3^2$
is 0.15~GeV$^2$. Note that the significant part of the error in the world 
average of $\alpha_s(M_Z)$ also come from the uncertainty due to
neglected HO QCD corrections. For this reason the two considered sources
of theoretical errors are correlated. Having no possibility to account 
for this correlation, we just combine both errors in quadrature and 
estimate the total theoretical error in $\Lambda_3^2$ as 0.16~GeV$^2$.
One can see that the uncertainty in $\Lambda_3$ is dominated by the 
experimental error, moreover accounting for the correlations of the 
separate sources of the theoretical error would decrease the latter.

{\bf 3.} We also extracted the HT contribution to $F_{2,3}^{\nu N}$ from 
the CCFR data of Ref.~\cite{Seligman:1997mc} using the approach
with $\alpha_s$ fixed.
The value of the parameter $\Lambda_3^2$ obtained from the 
NLO QCD analysis of those data with $x<0.7$
is given in  Table~1. The CCFR data are sensitive to the 
parameter $\Lambda_2$ too, although the precision is 
poor and the fitted value 
is comparable with zero within the errors\footnote{The CCFR data 
of Ref.~\protect\cite{Seligman:1997mc} are 
being revised now and for this reason the results 
obtained from the analysis of these data should be considered as preliminary.}.
The theoretical error in $\Lambda_3^2$
estimated in the same way as for the analysis of our 
data is $0.12~{\rm GeV}^2$. Results for both experiments are comparable 
within the errors and combining them we obtain the average 
\begin{equation}
\Lambda^2_3=0.44\pm0.19~({\rm exp})~{\rm GeV}^2. 
\label{eqn:lam}
\end{equation}

In order to check universality of the IRR model scales with respect 
to the specific choice of structure function
in the DIS process, we compared this value 
with the results of Ref.~\cite{Dasgupta:1996hh} on the analysis of the 
charged-leptons DIS data. 
Since in Ref.~\cite{Dasgupta:1996hh} the results are given in terms of 
parameter $A_2'$,
we transformed our average (\ref{eqn:lam}) using Eq.~(\ref{eqn:a2lam}). 
As a result, we obtain that for $n_f=3$ 
\begin{equation}
A_2'(F_3^{\nu N})=-0.130\pm0.056~({\rm exp})~{\rm GeV}^2.
\end{equation}
This value is smaller, than $A_2'(F_2^{e N})=-0.2$~GeV$^2$, 
given in Ref.~\cite{Dasgupta:1996hh}, 
although within the errors both values are comparable.
More precise conclusion about universality of the IRR model 
scales may be 
derived from the analysis of experimental data with improved statistics,
which have been collected using the IHEP-JINR Neutrino Detector 
with a different configuration of the neutrino beam channel.
The analysis of these data in order to extract the structure
functions is currently in progress. The data from the NuTeV collaboration  
\cite{NUTEV}, after their processing have been completed, may  
also be used to improve the precision   
of the IRR scales determination. In far sight 
a potential neutrino factory would allow for a detailed cross-check 
of the IRR model predictions and, in particular,
the determination of the IRR scales with an accuracy of several 
percent \cite{NUF}.

In conclusion, we extract the high-twist contribution to the 
neutrino-nucleon structure function $xF_3^{(\nu+\overline{\nu})N}$ 
from the analysis of the data collected in the first runs of the 
IHEP-JINR Neutrino Detector at the IHEP U-70 accelerator. 
We observe the negative HT contribution to the structure 
function $xF_3^{(\nu+\overline{\nu})N}$
which supports the earlier observations. 
The normalization scale of the IRR model 
extracted from the combined analysis of the IHEP-JINR and CCFR experiments 
is about $1\sigma$ lower than the one extracted from the 
data on the structure function $F_2^{lN}$ for the DIS of charged leptons.

{\bf Acknowledgments}

 We are indebted to V.Braun, A.Kataev, A.Sidorov,
E.Stein, and C.Weiss for stimulating discussions.
The work was supported by the RFBR grant 00-02-17432.

\end{document}